\documentclass[12pt, a4paper]{article}

\usepackage[top=1in, bottom=1in, left=1in, right=1in]{geometry}
\usepackage[T1]{fontenc}
\usepackage[table]{xcolor}
\usepackage{pgfplots}
\usepackage{cite}
\usepackage{amsmath,amssymb,amsfonts}
\usepackage{url}
\usepackage{algorithmic}
\usepackage{array}
\usepackage{float}
\usepackage{graphicx}
\usepackage{subcaption}
\usepackage{textcomp}
\usepackage{multirow}
\usepackage{multicol}
\usepackage{booktabs}
\usepackage{tabularx}

\title{\textbf{GNSS Spoofing Threat for V2X communications}}

\author{%
  Adolfo P. Jiménez$^{1,2}$,
  Juan Arquero-Gallego$^{1,2}$,
  Mario P. Luna$^{1,2}$,\\
  José E. Naranjo$^{1,2}$,
  Felipe Jimenez Alonso$^{1}$
  \thanks{$^{1}$Institute for Automobile Research (INSIA), Universidad Politécnica de Madrid, 28031 Madrid, Spain}
  \thanks{$^{2}$ETSI Sistemas Informáticos, Universidad Politécnica de Madrid, 28031 Madrid, Spain}
  \thanks{This paper is part of the project SAFE4CAR (grant number PID2022-140554OB-C31) funded by
    MCIN/AEI/10.13039/501100011033 and by FEDER, UE., and Comunidad de Madrid
    (SEGVAUTO‑5G‑CM, Ref.\ TEC‑2024/ECO‑277) and Strengthening C-ITS Adoption and
    Lining-up across Europe (SCALE) funded by CEF-T-2023-SIMOBGEN -- 101172496.}
  \thanks{Corresponding Author: Juan Arquero-Gallego, E-Mail: j.arquero@upm.es}
}

\date{}

\begin{document}

\maketitle

\begin{center}
\small\textit{%
\copyright~2026 IEEE\@. Personal use of this material is permitted.
Permission from IEEE must be obtained for all other uses, in any current or
future media, including reprinting/republishing this material for advertising
or promotional purposes, creating new collective works, for resale or
redistribution to servers or lists, or reuse of any copyrighted component of
this work in other works.}
\end{center}

\begin{abstract}
Global Navigation Satellite Systems (GNSS) constitute a core technology for delivering crucial positioning, navigation, and timing (PNT) services in the Vehicle-to-Everything (V2X) domain, where they are indispensable for generating Cooperative Awareness Messages (CAM) that uphold network reliability and vehicular safety. Yet, GNSS signals are acutely exposed to spoofing, an advanced attack in which an adversary transmits crafted signals that replicate legitimate satellite characteristics, misleading the receiver into computing a false position. This work presents a methodology for conducting physical spoofing with inexpensive Software Defined Radio (SDR), describing a coordinate generation pipeline that employs Haversine-based distance calculations, temporal discretization to emulate constant velocity, and linear interpolation to produce high-fidelity GPS baseband signals. The proposed attack is experimentally validated on real Commsignia OnBoard Unit (OBU) and RoadSide Unit (RSU) devices using a HackRF One across three scenarios that emulate synthetic trajectories at steady speeds of 90 km/h, 145 km/h, and 200 km/h. The most significant contribution of this paper is the demonstration that V2X communications are not secured, as they are susceptible to GNSS spoofing attacks, which cause service degradation without being detected.
\end{abstract}

\section{Introduction}
Global navigation satellite systems have become a fundamental infrastructure of modern society, forming the backbone of positioning, navigation, and timing services for critical infrastructures. It is estimated that any significant disruption to GNSS services could cause substantial economic losses due to their widespread use in logistics, transportation, and telecommunications \cite{NIST_Economic}. This strong dependence on GNSS-based positioning, navigation, and timing capabilities makes many sectors of the economy particularly vulnerable to service interruptions or signal manipulation.

Among these sectors, maritime transportation is highly dependent on GNSS for accurate vessel tracking, safe navigation along trade routes, and integration with electronic navigation systems such as electronic chart display and information systems (ECDIS), contributing to increased operational efficiency and maritime safety \cite{Zalewski2022}. Similarly, in the aviation sector, GNSS constitutes a technological pillar for performance-based navigation (PBN), enabling optimized flight routes and ensuring the safety and efficiency of air traffic operations \cite{AirTrafficGNSS2017}. Consequently, any interruption or degradation of GNSS services has a direct and immediate impact on civil and commercial aviation operations \cite{EASA2024}. These threats are particularly critical in emerging V2X applications, where accurate and reliable positioning is essential for cooperative driving, collision avoidance, and autonomous vehicle coordination.

The reliability of the GNSS signal can be compromised by various natural phenomena. Geomagnetic storms and disturbances in the ionosphere can cause phase errors and loss of tracking \cite{Strong_Geomatic_Storms,Spring_Multipath}, severely affecting the accuracy of the receiver. Similarly, the effect of \textit{multipath} or signal bounce from physical structures generates significant delays that degrade position calculation \cite{IEEE_Multipath}. In addition to these natural degradations, there are intentional threats: jamming and spoofing. This consists of the intentional emission of radio frequency (RF) interference to overpower or disrupt authentic GNSS signals \cite{Medina2019MaritimeJamming}.
In contrast, spoofing is a more sophisticated technique that emits false signals that mimic the structure of authentic satellite signals, with the aim of manipulating the location reported by the receiver \cite{Humphreys2012}.

During a spoofing attack, the attacker emits a synthetic replica of a legitimate signal with greater power, impersonating the signal, and deceiving the receivers. This process requires a strategic physical arrangement that allows the false wavefront to override the authentic one at the receiver's front end without interrupting service continuity. Although these attacks formerly required expensive professional-grade simulators, the rise of SDR has made this vulnerability more feasible, allowing the generation of false GNSS signals using low-cost hardware \cite{Humphreys2008Spoofer}.

In the specific context of V2X communications, GNSS is necessary to generate CAM messages. A successful attack in this ecosystem seeks to compromise the integrity of the network, which could have serious consequences, such as generating false collision alerts or manipulating traffic flows. To apply this attack to a V2X system, it is necessary to generate a false GPS signal that will be acquired by the OBU, which will use it to calculate different variables and provide erroneous data to the RSU.

Even though advanced countermeasures exist in the wider GNSS domain, such as Galileo's OSNMA authentication service \cite{OSNMA_Galileo}, their deployment within the V2X application domain is still at a very early stage. Currently, commercial V2X architectures lack robust protection mechanisms deployed on a large scale, leaving intelligent transport systems (ITS) vulnerable to manipulation by their main location provider. The main contribution of this work is the experimental validation of spoofing attacks on physical V2X infrastructure using low-cost SDR hardware. Unlike previous simulation-based studies, this paper validates the technical feasibility of compromising RSU and OBU data integrity by executing GNSS spoofing via HackRF One.

\section{Related Work}
\label{sec:related_work}

V2X security has been primarily studied at the network and application layers, often assuming the integrity of GNSS positioning data and relying almost exclusively on simulated environments. Gularte et al. \cite{Gularte2024V2XSecuritySurvey} and Wellens-Miles et al. \cite{WellensMilesSLR} provide comprehensive taxonomies of the V2X cybersecurity landscape; however, their analyzes are limited to literature reviews and theoretical threat models. Similarly, Hasan et al. \cite{Hasan2020SecuringV2X} offer an extensive overview of security challenges and countermeasures for V2X communication platforms. Similarly, Tzoannos et al. \cite{TzoannosSpoofing} investigate the impact of GNSS spoofing on traffic flow, but their work relies on simulated data injection to evaluate macroscopic congestion metrics, without assessing the feasibility of such attacks on real hardware.  In contrast to these theoretical and simulation-based approaches, recent studies have begun to address the physical layer. Sousa et al. \cite{Sousa2025JammingV2X} experimentally evaluate \textit{jamming} attacks that affect the DSRC and C-V2X sidelinks; however, their work focuses on denial-of-service interference rather than signal deception and does not consider GNSS manipulation. Krayani et al. \cite{Krayani2023V2XSpoofingJamming} propose theoretical countermeasures against spoofing, but without experimental validation on real vehicular communication hardware. Sanders and Wang \cite{SandersWang2020V2XSpoofing} study the location of GNSS spoofing attacks using V2V communications, but their analysis is based on simulations and does not evaluate the impact of spoofing signals on OBU--RSU communication links. The existing literature shows a gap in experimental methods for executing GNSS spoofing in real V2X systems. This study uses SDR to demonstrate how spoofed signals compromise the OBU--RSU link and disrupt RSU data processing, providing an assessment of physical-layer vulnerabilities under real conditions.
Table~\ref{table:comparative} provides a comparative overview of key studies.

\begin{table}[htbp]
\centering
\scriptsize
\begin{tabularx}{\textwidth}{|l|X|X|}
\hline
\textbf{Ref.} & \textbf{Research Focus} & \textbf{Difference with this Work} \\ \hline
\cite{WellensMilesSLR}, \cite{Gularte2024V2XSecuritySurvey} & Systematic literature reviews on V2X cybersecurity. & Documentary analysis; lacks empirical or experimental hardware validation. \\ \hline
\cite{Hasan2020SecuringV2X} & Analysis of security threats and countermeasures in V2X communication platforms. & Theoretical and survey-based; lacks experimental evaluation of GNSS spoofing. \\ \hline
\cite{TzoannosSpoofing} & Simulation of spoofing impact on global traffic flows. & Focuses on congestion metrics; ignores real hardware front-end response. \\ \hline
\cite{HuangFalsified} & Logical modeling of data injection in BSM messages. & Operates at the application layer, omitting physical RF vulnerabilities. \\ \hline
\cite{AnsariVASP} & Framework for V2X message and protocol emulation. & Software-controlled environment; no interaction with physical sensors. \\ \hline
\cite{Sousa2025JammingV2X} & Experimental jamming attack on DSRC/C-V2X sidelinks. & Focuses on signal interference (DoS), not on coordinate deception (Spoofing). \\ \hline
\cite{Krayani2023V2XSpoofingJamming} & Design of theoretical algorithms for signal detection. & Defensive approach; lacks validation against real-world SDR attackers. \\ \hline
\cite{SandersWang2020V2XSpoofing} & Localizing GNSS spoofing attacks on vehicular GPS using Vehicle-to-Vehicle (V2V) communications, focusing on theoretical modeling and simulation of attack localization. & Focuses on theoretical V2V localization; does not evaluate the impact of the attack on OBU--RSU communication links. \\ \hline
\textbf{Prop.} & \textbf{Experimental GNSS spoofing attacks targeting V2X communications; SDR-based evaluation of OBU--RSU link compromise.} & --- \\ \hline
\end{tabularx}
\caption{Comparison between related work and the proposed approach.}
\label{table:comparative}
\end{table}

\section{Theoretical Basis}

\subsection{GPS L1 C/A Signal Characterization}

To perform a GNSS spoofing attack, a counterfeit signal must replicate the structure
of a genuine transmission. In this work, the target is the GPS L1 C/A signal, which is transmitted at a carrier frequency $f_{L1}=1575.42~\text{MHz}$ and consists of a radio-frequency carrier, a pseudorandom noise (PRN) spreading code and a navigation data message. Signal separation among satellites is achieved through Code Division Multiple Access (CDMA).
Each satellite $k$ is assigned a unique PRN sequence $c_k[n]$ of length 1023 chips,
generated as a Gold code by mod 2 addition of two maximal-length sequences produced
by linear feedback shift registers (LFSRs), namely $G_1$ and $G_2$ \cite{Kaplan2005}:
\begin{equation}
c_k[n] = G_1[n] \oplus G_2[n-\tau_k],
\end{equation}
where $\tau_k$ denotes a delay in a satellite-specific code.

\begin{figure}[htbp]
    \centering
    \includegraphics[width=0.7\textwidth]{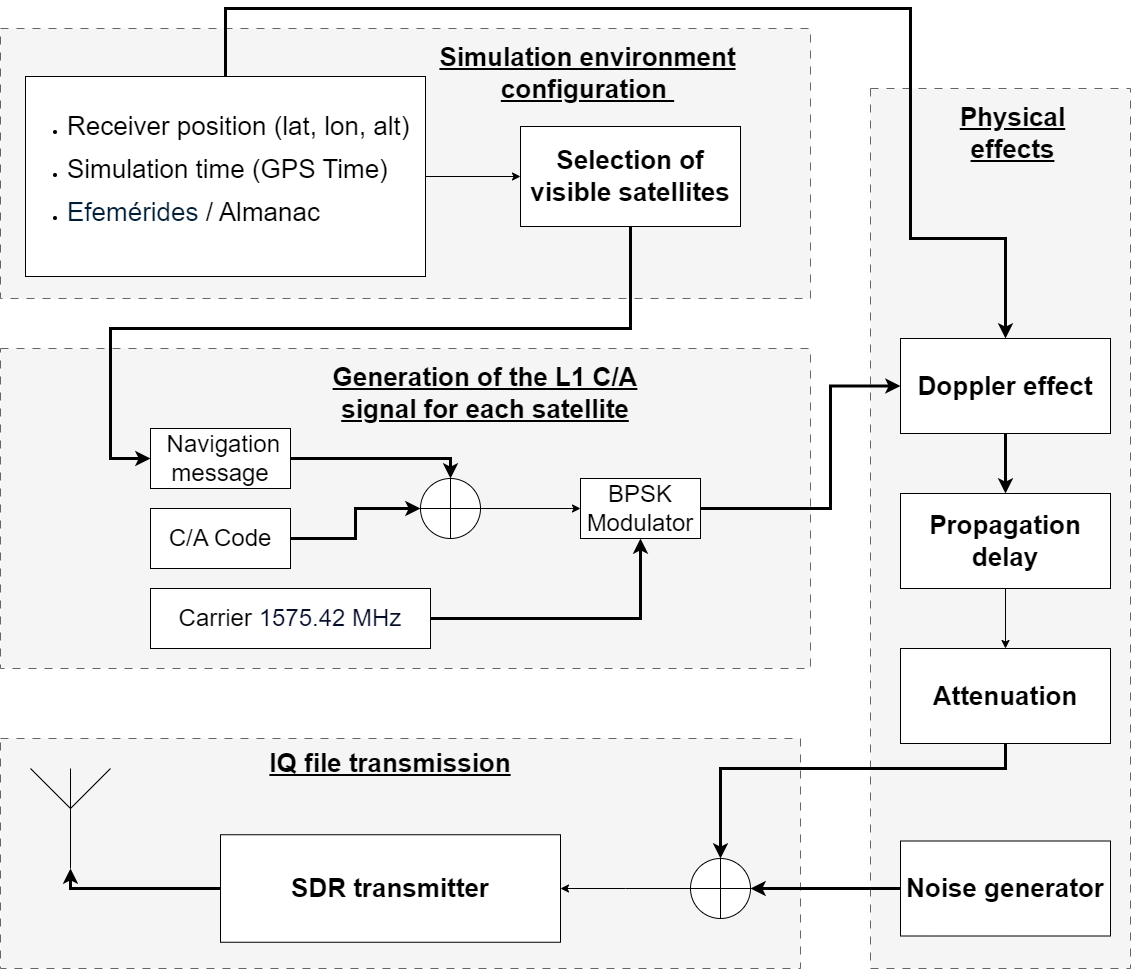}
    \caption{Functional block diagram of the GPS L1 C/A signal simulation and transmission pipeline.}
    \label{fig:simgps}
\end{figure}

The legacy civil GPS L1 C/A signal employs Binary Phase Shift Keying (BPSK) modulation.
In BPSK, the carrier phase is shifted by $0^\circ$ or $180^\circ$ according to the sign
of the baseband signal. The passband signal transmitted by the $k$-th satellite can therefore
be modeled as:
\begin{equation}
s_k(t) = \sqrt{2P}\, c_k(t)\, d_k(t)\cos\!\left(2\pi f_{L1} t\right),
\end{equation}
where $s_k(t)$ denotes the transmitted signal, $P$ is the signal power, $c_k(t)$ is the PRN
spreading code, $d_k(t)$ represents the navigation data symbols, and $f_{L1}$ is the L1
carrier frequency.
Figure~\ref{fig:simgps} shows the simulation chain of the GPS L1 C/A signal from PRN and data generation to the complex formation of IQ samples for the transmission of SDR \cite{Kaplan2005,Benson2006,Majoral2024}.

\subsection{Mathematical Modeling of the Jamming--Spoofing Attack}

Combined \emph{jamming} and \emph{spoofing} attacks on the GPS L1 signal exploit the internal tracking loops of the receiver. These loops maintain synchronization with the satellite signal: DLL (\emph{Delay Lock Loop}) tracks the PRN code (Pseudo-Random Noise) to accurately estimate the signal's time of arrival, while PLL (\emph{Phase Lock Loop}) maintains carrier phase coherence, which is fundamental to GNSS receiver operation and its vulnerability to interference \cite{Kaplan2005}.

The initial effect of jamming can be modeled as a degradation of the receiver's effective signal-to-noise ratio due to the added interference power, which reduces the apparent carrier-to-noise density ratio (\(C/N_0\)). This model captures how jamming distorts the correlator response and tracking performance \cite{Psiaki2016,Rados2024}:

\begin{equation}
(C/N_0)_{\text{eff}} = \frac{C}{N_0 + J \cdot BW_{ss}}
\end{equation}

where \(C\) is the power of the authentic GNSS signal, \(N_0\) the noise spectral density, \(J\) the interference power, and \(BW_{ss}\) the spreading bandwidth of the signal. The increase in \(J \cdot BW_{ss}\) reduces the perceived \(C/N_0\) at the receiver, increasing the variance of DLL error and weakening the sharpness of the PRN code correlation peak. As jamming degrades tracking, a spoofing signal with a slightly stronger correlation peak can dominate the receiver's correlator during reacquisition \cite{Psiaki2016,Rados2024}.

When the receiver attempts to re-lock the carrier, the PLL discriminator observes the vectorial combination of the authentic and spoofing signals. The resulting phase error, which determines whether the PLL locks onto to the authentic or impostor signal, can be analytically expressed as \cite{Psiaki2016,Ghizzo2025}:

\begin{equation}
\phi_{err} = \operatorname{atan2}\big(Q_a + \sqrt{P_s}\sin\Delta\phi,\; I_a + \sqrt{P_s}\cos\Delta\phi\big)
\end{equation}

Here, \(I_a\) and \(Q_a\) are the in-phase and quadrature components of the authentic signal at the receiver's correlator. The spoofing signal can be represented as a complex phasor:

\begin{equation}
I_s + j Q_s = \sqrt{P_s}\cos\Delta\phi + j \sqrt{P_s}\sin\Delta\phi = \sqrt{P_s}\, e^{j\Delta\phi}
\end{equation}

where \(P_s\) is the linear power of the spoofing signal and \(\Delta\phi\) its intentionally introduced phase offset relative to the authentic signal. The total phase error measured by the receiver is then determined by the sum of the vector components of authentic and spoofing signals in the I/Q plane. If the spoofing signal has a magnitude comparable to or greater than the authentic signal, the resulting phase \(\phi_{err}\) will tend toward the spoofing signal's phase, and the PLL will lock onto the false signal \cite{Ghizzo2025}.

\subsection{Power-Domain Feasibility Analysis}
The technical feasibility of a spoofing attack against an OBU using an SDR platform such as HackRF One is based on the enormous power imbalance between legitimate GNSS signals and generated signals.
To represent the intensity with which the false signal would reach the OBU ($P_r$), a comprehensive link budget analysis has been performed to characterize how the signal evolves as a function of distance and random channel conditions.

\begin{equation}
P_r = P_t + G_t + G_r - L_{tx} - L_{rx} - FSPL
\end{equation}

The received power is determined by the transmitted power $P_t$ expressed in dBm, which denotes power on a logarithmic scale referenced to 1 mW. The antenna gains $G_t$ and $G_r$ are given in dBi, specifying the gain of each antenna relative to an ideal isotropic radiator. The terms $L_{tx}$ and $L_{rx}$ are expressed in dB, as they represent relative attenuation due to system losses: $L_{tx}$ accounts for feeder and connector losses on the transmitter side, while $L_{rx}$ models front-end and cable losses at the receiver. In addition, the free-space path loss (FSPL) \cite{RappaportWireless} is included to capture the attenuation undergone by an electromagnetic wave when it propagates without obstructions. It is given by:

\begin{equation}
FSPL(d, f) = 20 \log_{10}(d) + 20 \log_{10}(f) +
20 \log_{10}\left(\frac{4\pi}{c}\right)
\end{equation}

where \(d\) is the separation between transmitter and receiver, \(f\) is the carrier frequency, and \(c\) is the speed of light in vacuum. This logarithmic formulation expresses the loss in decibels and highlights that attenuation grows with both distance and frequency. Because the analysis is conducted in the logarithmic domain, all elements of the link budget can be combined through straightforward addition and subtraction.

To approximate realistic propagation conditions, the model also includes shadowing $L_{shadow}$ and Rayleigh fading $L_{fading}$. Shadowing captures large-scale, slowly varying changes in received signal power caused by obstacles in the environment. In the logarithmic (dB) domain, it is modeled as a Gaussian random variable with zero mean and standard deviation $\sigma_{shadow}$, which governs the intensity of these slow power fluctuations.

Conversely, Rayleigh fading characterizes rapid small-scale fluctuations due to multipath propagation. It is described in the linear domain by a Rayleigh-distributed random variable that models the signal envelope when no dominant line-of-sight component is present. The resulting amplitude is then transformed into the logarithmic (dB) domain and incorporated into the overall link budget \cite{RappaportWireless}.

\begin{equation}
P_r = P_r - L_{shadow} - L_{fading}
\end{equation}

To assess the generated spoofing signal against the genuine GPS signal, the signal-to-noise ratio (SNR) of the spoofing signal at the receiver is compared with that of the authentic GPS signal, which is approximately -21 dB \cite{InsideGNSS2010SignalStrength}.

\begin{equation}
SNR = P_r - (-174 + 10 \log_{10}(B) + NF)
\end{equation}

\begin{figure}[htbp]
    \centering
    \includegraphics[width=0.8\textwidth]{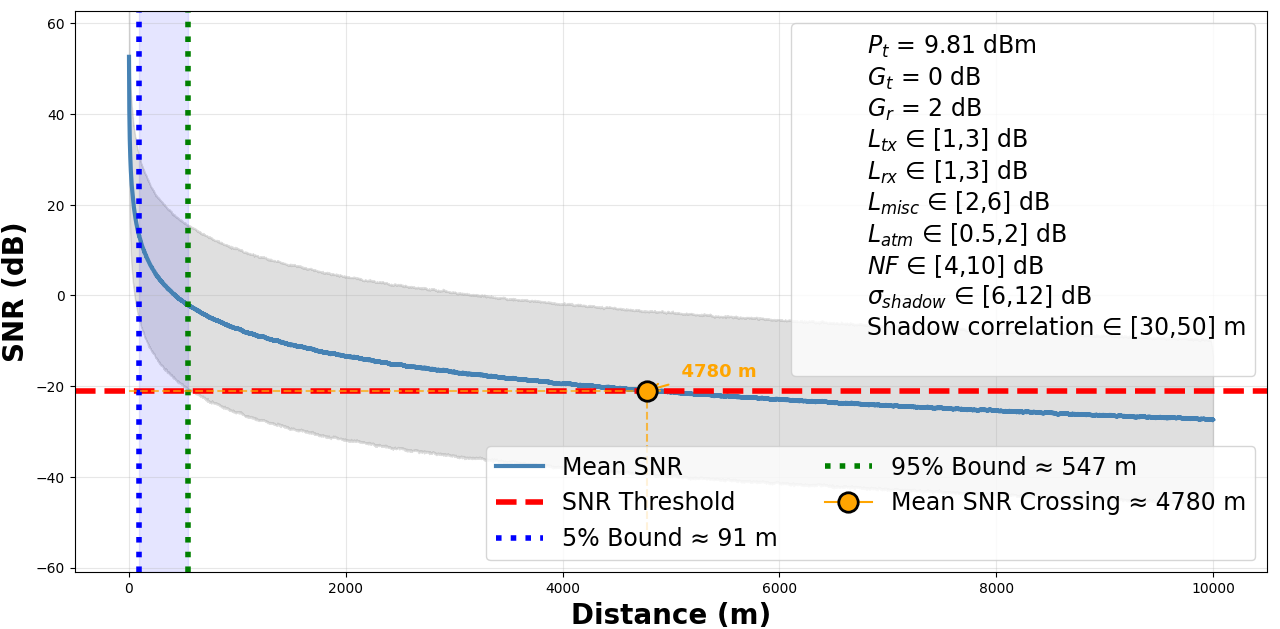}
    \caption{SNR vs.\ distance. Monte Carlo analysis (30,000 runs).}
    \label{fig:trayectoria_montecarlo}
\end{figure}

Here, $-174$ dBm/Hz denotes the thermal noise spectral density at room temperature, $B$ is the receiver bandwidth in hertz and $NF$ is the receiver noise figure in decibels.
To account for the impact of random effects, the model was executed 30,000 times using a Monte Carlo simulation that included both shadowing and fading. For each realization, the distance was recorded at which condition $SNR \leq SNR_{gps}$ was first satisfied, identifying the point beyond which the spoof attempt could no longer succeed. The resulting statistical distribution is presented in Figure~\ref{fig:trayectoria_montecarlo}, which displays the 5--95\% confidence interval, representative crossing trajectories, and their intersections with the threshold. Based on this analysis, the effective spoofing range lies between roughly 91 and 547 meters. Beyond 4.8 km, the performance degrades gradually rather than suddenly: at this range, roughly half of the simulated scenarios have already lost the link, while the other half still maintain it. From 10 km onward, nearly all realizations fall below the threshold, rendering successful spoofing extremely unlikely.

\section{Methodology}

\subsection{Controlled Testbench for GNSS Spoofing Evaluation}
To assess the effects of GNSS spoofing on vehicular navigation systems, a controlled experimental setup is defined in which a test vehicle equipped with an OBU drives along a road while receiving GNSS signals. Under nominal conditions, the OBU determines its position based on genuine satellite transmissions.

This study focuses on a \emph{carry-off GNSS spoofing} attack, where counterfeit signals are first synchronized with legitimate GNSS signals and then gradually amplified in power until the receiver begins to track them. In this way, carefully controlled positioning errors are introduced without causing an abrupt loss of signal, making the attack difficult to detect.

Within the CAPEC (Common Attack Pattern Enumeration and Classification) framework~\cite{CAPEC628}, a carry-off GNSS spoofing attack is classified as a type of attack that is highly unlikely to occur due to the knowledge and materials required to carry it out, but if carried out effectively, its consequences can have a major impact on systems. However, with the advent of SDR technology, the feasibility of such attacks has significantly increased, transitioning from a theoretical risk to a practical concern.

In the scenario considered, an SDR-based transmitter is installed near the road infrastructure to broadcast spoofed GNSS signals in the same frequency band as legitimate satellite signals. Consequently, the OBU simultaneously receives authentic and forged signals, allowing analysis of receiver behavior and measurement of positioning errors induced by the carry-off spoofing attack.

\begin{figure}[htbp]
    \centering
    \includegraphics[width=0.6\textwidth]{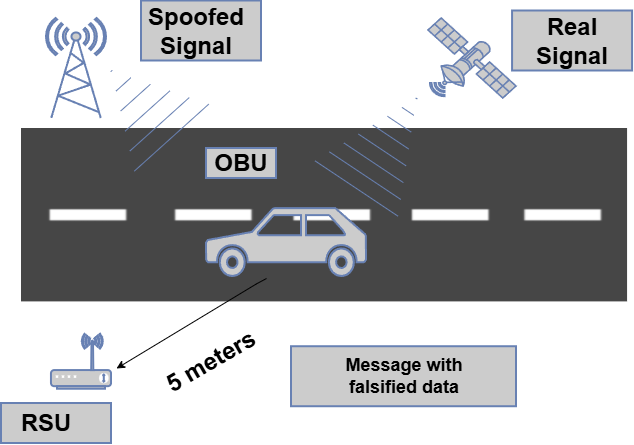}
    \caption{Evaluation testbench for GNSS spoofing attacks in V2X networks.}
    \label{fig:spoofing_scenario}
\end{figure}

Figure~\ref{fig:spoofing_scenario} illustrates the conceptual scenario of GNSS spoofing, including vehicle, genuine satellite signals, and the source of SDR-based spoofing.

The experimental assessment is performed on a controlled testbench that integrates commercial vehicular communication devices with an SDR-based platform. The vehicle hosts a Commsignia OBU featuring an embedded GNSS receiver, while counterfeit GPS L1 signals are produced using GPS-SDR-SIM and broadcast over the air through GNU Radio and a HackRF SDR.

\subsection{Developed Methods}
To generate realistic GPS signals, a routing engine provides a discrete set of road coordinates that define the vehicle trajectory. These coordinates must be processed to ensure that the simulated receiver follows a smooth path at constant velocity, with uniform temporal sampling. To this end, the proposed method combines great-circle distance computation, constant-velocity kinematics, uniform time discretization, and linear interpolation of geographic coordinates.

\subsubsection{Distance Computation and Time Discretization}
The separation between consecutive route nodes must account for the curvature of the Earth. Consequently, the distance between two geographic coordinates $(\varphi_1,\lambda_1)$ and $(\varphi_2,\lambda_2)$ is calculated using the Haversine formula, which provides a numerically stable estimate of the great-circle distances. The central angle between the two points is given by
\begin{equation}
a = \sin^2\!\left(\frac{\Delta \varphi}{2}\right) + \cos(\varphi_1)\cos(\varphi_2)\sin^2\!\left(\frac{\Delta \lambda}{2}\right)
\end{equation}
and the corresponding surface distance is obtained as
\begin{equation}
d = 2R\,\operatorname{atan2}\!\left(\sqrt{a},\sqrt{1-a}\right)
\end{equation}
where $R = 6371$~km denotes the mean Earth radius. The distance is converted to meters as $d_m = 1000d$.

Taking into account constant velocity along each segment of the route, the target speed $v_{kmh}$ is converted to meters per second, and the traversal time is calculated as $T = d_m / v$. The motion is discretized using a fixed sampling interval $\Delta t = 0.1$~s, which yields a total number of samples
\begin{equation}
N = \max\left(1,\left\lfloor \frac{T}{\Delta t} \right\rfloor\right)
\end{equation}
which ensures uniform temporal spacing while guaranteeing at least one sample per segment.

\subsubsection{Linear Interpolation of Coordinates}
Intermediate receiver positions are synthesized through linear interpolation between consecutive route nodes. For each discrete sample index $j \in [0,N]$, a normalized interpolation factor is defined as
\begin{equation}
f_j = \frac{j}{N}
\end{equation}

The interpolated latitude, longitude, and timestamp are computed as
\begin{align}
\varphi_j &= \varphi_1 + f_j(\varphi_2 - \varphi_1), \\
\lambda_j &= \lambda_1 + f_j(\lambda_2 - \lambda_1), \\
t_j &= t_0 + j \Delta t,
\end{align}
where $t_0$ denotes the cumulative time elapsed from previous segments. This interpolation strategy, combined with uniform temporal discretization, ensures a constant receiver velocity between nodes and produces smooth trajectories suitable for realistic GNSS signal generation.

\section{Test and Results}
\begin{figure}[htbp]
    \centering
    \includegraphics[width=0.7\textwidth]{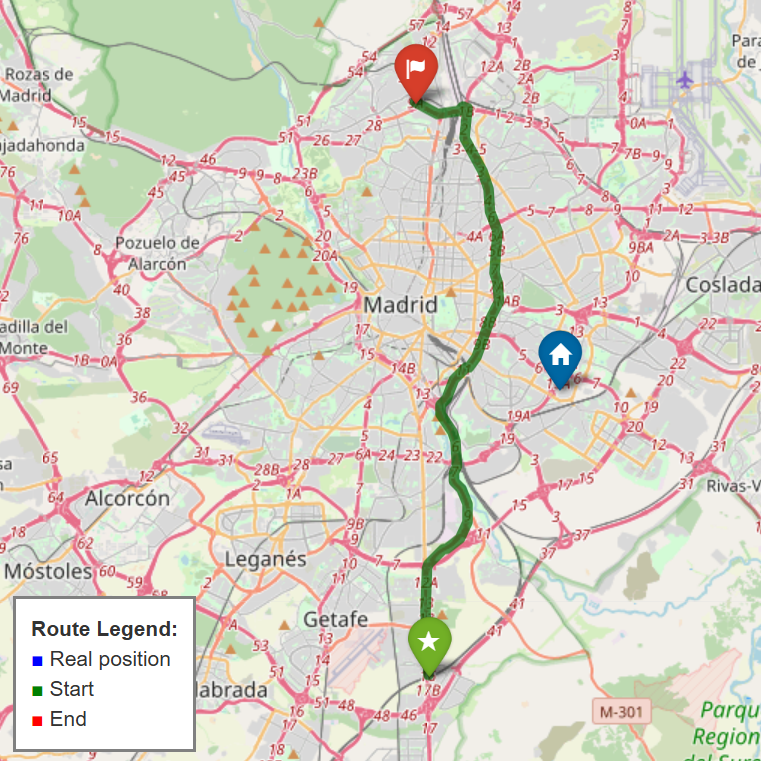}
    \caption{Vehicle trajectory after spoofing attack.}
    \label{fig:trayectoria}
\end{figure}

Three different dynamic scenarios were defined to evaluate how OBUs respond when a spoofed GPS position indicates motion at different speeds. In the first scenario (90 km/h), the goal is to make the system assume that it is operating under low-speed dynamic conditions, with relatively smooth and steady motion. The second scenario (145 km/h) evaluates an intermediate speed range, reflecting a more dynamic yet still stable driving state, where the vehicle travels at high but nearly constant speeds. Lastly, the third scenario (200 km/h) corresponds to a high speed regime, in which the vehicle dynamics become considerably more demanding, with very high speeds and pronounced acceleration peaks. All scenarios were carried out along the same route as depicted in Figure~\ref{fig:trayectoria}.

\begin{figure}[htbp]
    \centering
    \begin{subfigure}{0.32\textwidth}
        \centering
        \includegraphics[width=\linewidth]{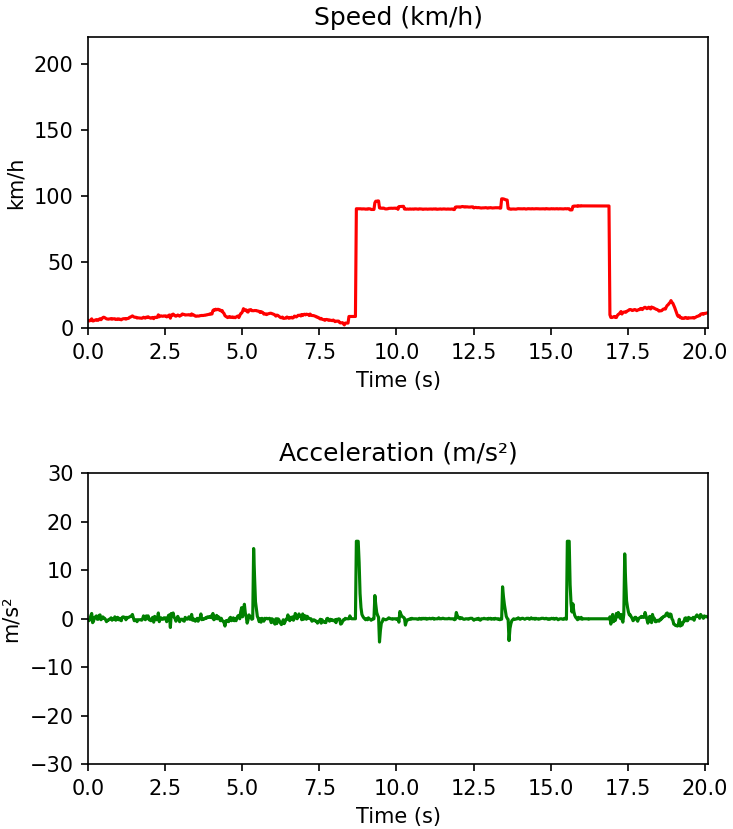}
        \caption{Scenario 1: 90 km/h}
        \label{fig:sc2_data}
    \end{subfigure}
    \hfill
    \begin{subfigure}{0.32\textwidth}
        \centering
        \includegraphics[width=\linewidth]{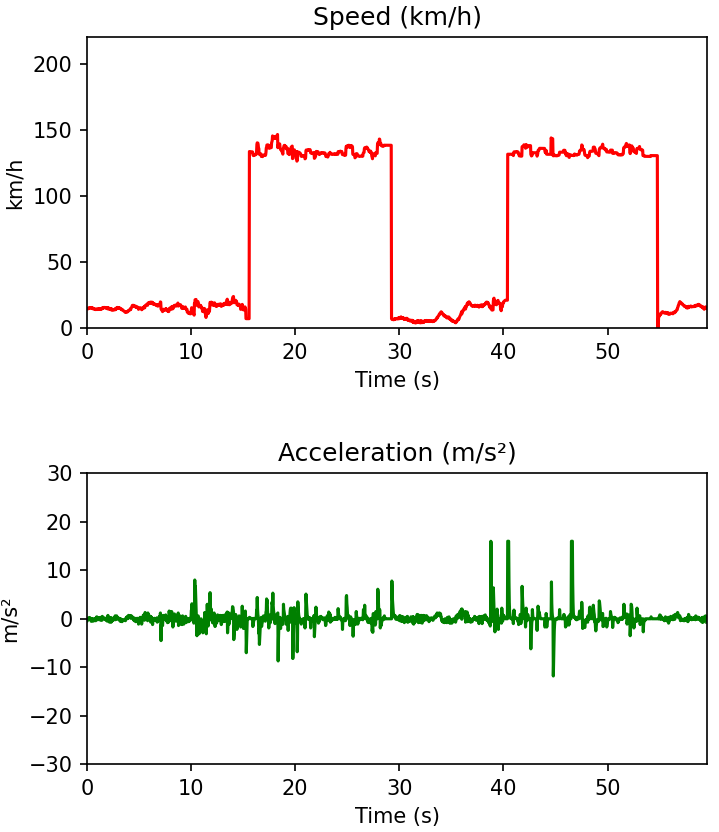}
        \caption{Scenario 2: 145 km/h}
        \label{fig:sc1_data}
    \end{subfigure}
    \hfill
    \begin{subfigure}{0.32\textwidth}
        \centering
        \includegraphics[width=\linewidth]{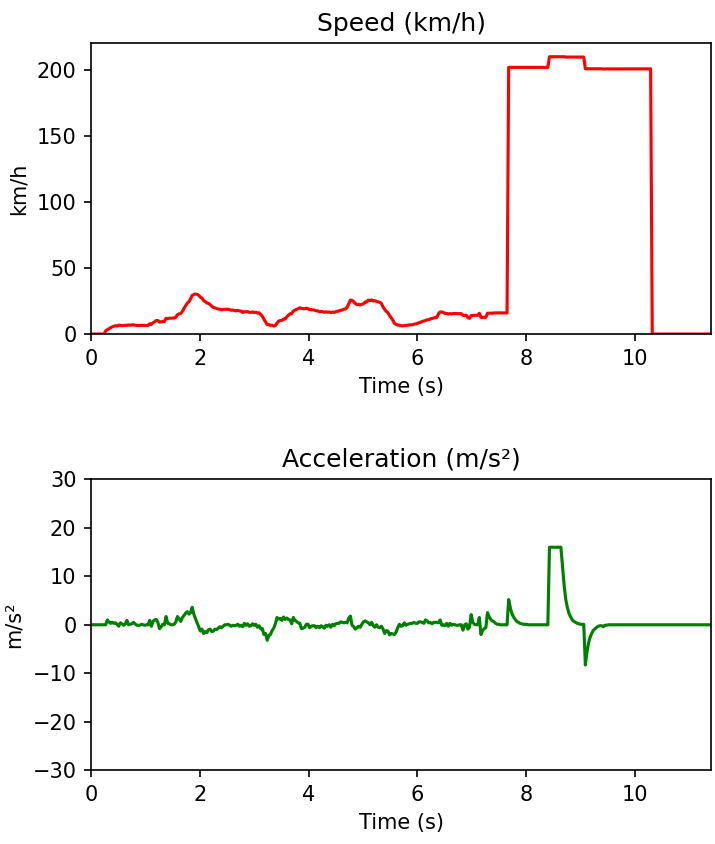}
        \caption{Scenario 3: 200 km/h}
        \label{fig:sc3_data}
    \end{subfigure}
    \caption{Experimental GNSS spoofing scenarios showing vehicle speed and acceleration profiles for three different velocities.}
    \label{fig:spoofing_scenarios}
\end{figure}

The resulting kinematic behavior for each scenario is shown in Fig.~\ref{fig:spoofing_scenarios}, which illustrates the speed and longitudinal acceleration profiles, corresponding to the vehicle dynamics fields defined in ETSI ITS-G5 \cite{ETSIEN3026372}. In all cases, two clearly distinct phases are visible: an initial normal-driving phase with smooth, moderate dynamics, followed by a spoofed phase in which tampered GNSS signals impose artificial trajectories.

\begin{figure}[htbp]
    \centering
    \includegraphics[width=0.85\textwidth]{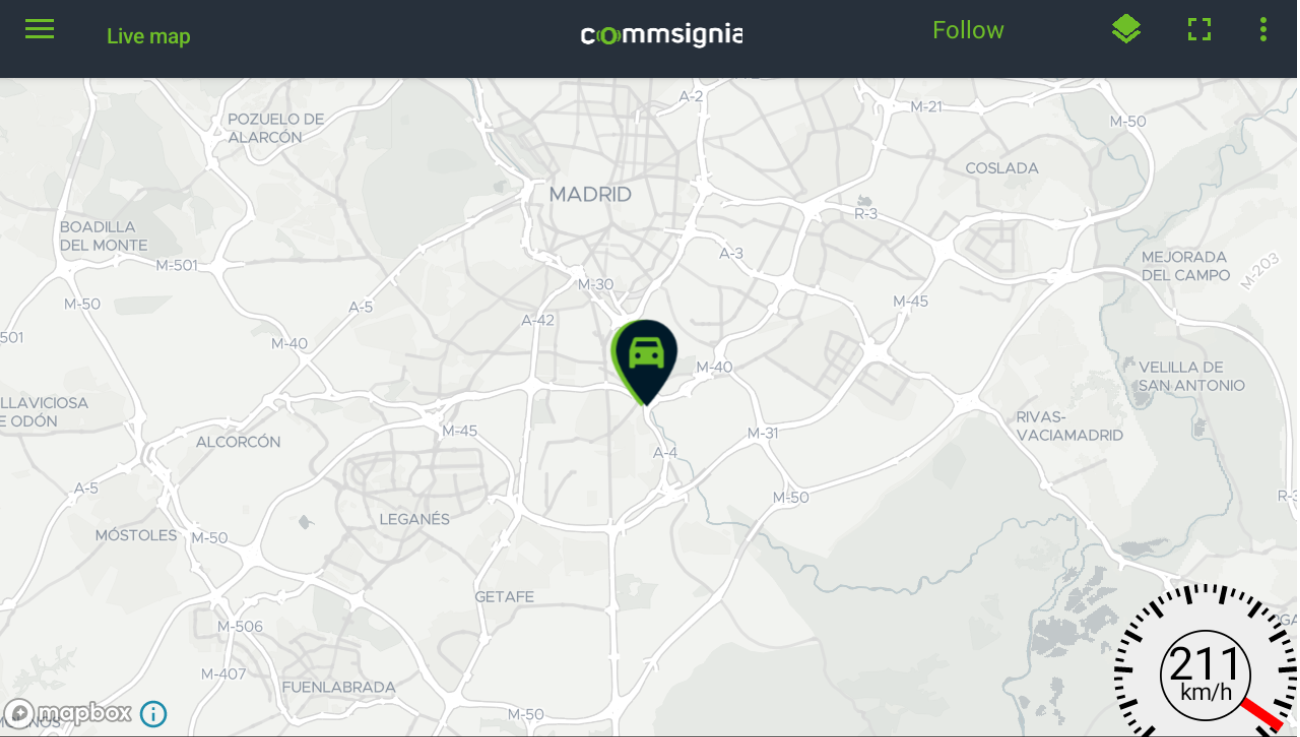}
    \caption{Real-time vehicle telemetry ($200 \text{ km/h}$) displayed on the Commsignia platform after RSU reading.}
    \label{fig:OBU200}
\end{figure}

These transitions occur without the usual acceleration and braking phases that a real vehicle would exhibit, flagrantly violating the basic laws of inertia and vehicle dynamics. Equally notable are the acceleration peaks, which reach extreme values close to 20 m/s². This level of acceleration exceeds the capabilities of standard vehicles, clearly indicating that the CAM messages have been manipulated with the intention of deceiving the V2X infrastructure. Furthermore, when the geographical context is taken into account, it becomes clear that the location communicated by the vehicle has been falsified: it is supposedly traveling through Madrid at 200 km/h on a stretch of highway, which is incompatible with the maximum speed limit.
The experimental results validate the success of the attack, as illustrated in Figure~\ref{fig:OBU200}: both the GPS coordinates and the speed have been effectively falsified.

\section{Conclusions and Future Work}
The results confirm that spoofing attacks against V2X infrastructure are physically feasible under realistic operating conditions. Both analytical modeling and experimental evaluation demonstrate that GNSS-based synchronization mechanisms used in V2X systems can be intentionally manipulated by controlled signal injection. As illustrated by the results obtained, these attacks are feasible under different test conditions, demonstrating that the V2X communication system is not protected against GNSS spoofing attacks. The data collected by real RSU stations contained in this paper represent manipulated data that has been intentionally introduced to cause service degradation and, ultimately, denial of service. It also demonstrates that these attacks can be carried out without being detected by the system, posing a real cybersecurity threat associated with the deployment of V2X technology.
These findings highlight the tangible and practical nature of the threat rather than a purely theoretical vulnerability. In addition, a practical field testing methodology has been developed to experimentally evaluate the vulnerability of the system and support the improvement and hardening of the V2X equipment against such attacks.
A future line of research is to study the behavior of RSU stations when connected to multiple OBUs, where only some of them send data based on the falsified signal, mixing the reception of intact signals and manipulated signals.

\end{document}